\documentclass[lettersize,journal]{IEEEtran}
\usepackage{amsmath,amsfonts}
\usepackage{algorithmic}
\usepackage{algorithm}
\usepackage{array}
\usepackage[caption=false,font=normalsize,labelfont=sf,textfont=sf]{subfig}
\usepackage{textcomp}
\usepackage{stfloats}
\usepackage{url}
\usepackage{verbatim}
\usepackage{graphicx}
\usepackage{cite}
\usepackage{listings}
\begin{document}

\title{Technical Aspects of Plasma Operational Simulation (POPSIM): A Framework for Data-Driven Simulation and Control}

\author{Allen M. Wang, Zander Keith, Mark Dan Boyer, Andrew Oakleigh Nelson, Alex Saperstein, Alessandro Pau, and Cristina Rea

\thanks{Allen M. Wang, Zander Keith, Alex Saperstein, and Cristina Rea are with the MIT Plasma Science and Fusion Center. Mark Dan Boyer is with Commonwealth Fusion Systems. Andrew Oak Nelson is with Columbia University. Alessandro Pau is with the Swiss Plasma Center at École Polytechnique Fédérale de Lausanne (EPFL).}
}

\markboth{To Be Submitted to IEEE Transactions on Plasma Science}%
{Shell \MakeLowercase{\textit{et al.}}: A Sample Article Using IEEEtran.cls for IEEE Journals}

\maketitle

\begin{abstract}
This paper reports on technical aspects of Plasma Operational Simulation (POPSIM), a research framework for data-driven simulation and control built in the machine learning framework JAX. The objective of the project is to address the extremely challenging simulation and modeling requirements of tokamak operations and control by combining simple principles-based models with data-driven models, spanning everything from power laws to new neural network architectures. This paper reports on key software and operations problems that the framework addresses, with examples from ongoing modeling activities for illustration.
\end{abstract}


\section{Introduction}
The control and operation of tokamaks places challenging requirements on simulation capabilities. The most immediate of these is the need for speed, driven by the timescales involved. For most contemporary tokamak experiments, there are only order 10 minutes between pulses, and the processing of critical data can take a majority of this time. Thus simulation models that are used to check control system changes need to complete full-shot simulations within $\mathord{\sim}1$ minute to allow for additional iteration and decision making time. Due to inevitable uncertainties and off-normal events, parallel simulations would ideally be performed across a wide distribution of outcomes to test for robustness. Further complicating the challenge, the application of trajectory and control optimization techniques, such as optimal control and reinforcement learning, would require $\mathord{\sim}10^1$ to $\mathord{\sim}10^3$ of sequential simulations. In addition, it is desirable to have real-time simulators run as part of the plasma control system (PCS) to inform control decisions. Such simulators need to run on timescales of a millisecond and have an extremely high degree of numerical stability and reliability; a simulator running at 1kHz would be called $\mathord{\sim}10^4$ times over the course of a SPARC primary reference discharge (PRD) pulse \cite{creely2020overview}.

Another set of requirements is driven by the practical realities of tokamak control and operations which challenges the application of many physics simulation codes. To summarize some key considerations:
\begin{itemize}
    \item The plasma state is only partially observable given the available diagnostic suites.
    \item Many variables used as boundary conditions in physics simulation, such as edge density and temperature, are uncontrollable in practice.
    \item Models of hardware dynamics, power supplies for example, typically need to be inferred from experimental data and can change over time.
\end{itemize}
These issues motivate the development of data-driven simulators that predict the time evolution of quantities observable by the diagnostic suite, ideally given only controllable variables as inputs. If additional uncontrollable variables are necessary, the simulator should be parallelized over a range of possible values. The simulators should, in addition, be continuously tunable to experimental data and validated at scale.
\begin{figure}[!t]
    \centering
    \includegraphics[width=\linewidth]{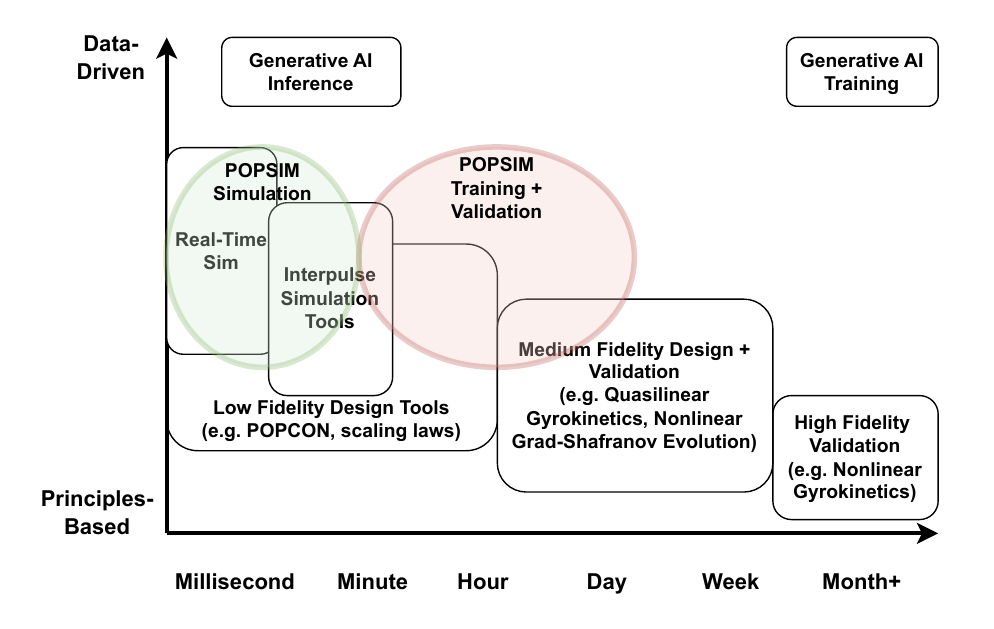}
    \caption{Qualitative depiction of classes of simulations in the space of simulation time and data-driven vs. principles-based. The simulation space that the POPSIM framework aims to support is shown in green, along with the timescales targeted for training and validation of simulation models.}
    \label{fig:sim-space}
\end{figure}

Such a set of requirements is not entirely unique. For example, a similar set is present for the problem of aircraft control design. In that context, classical \textit{system identification} techniques are typically employed to derive models of aircraft dynamics from flight data \cite{chyczewski2022summary, morelli2016aircraft}. Indeed, classical system identification techniques have been employed successfully to certain plasma control tasks such as detachment control \cite{ravensbergen2021real}. However, classical system identification techniques are typically restricted to linear models, and it is clear that tokamak plasmas exhibit many nonlinear effects that should be captured. Fortunately, the advent of new machine learning frameworks, such as JAX \cite{jax2018github}, that enable the easy expression of models which combine physics equations with neural networks present new opportunities for pushing system identification much further. This new paradigm is often referred to as \textit{Scientific Machine Learning} (SciML). Recent experiments at the Tokamak à Configuration Variable (TCV) facility have demonstrated that the SciML approach can yield predictive time-dependent models with a modest amount of training data \cite{wang2025learning}. Furthermore, the utility of these models was demonstrated by performing rampdown trajectory optimization on them, which resulted in a reduction in the disruptivity during the ramp-down of the plasma current \cite{wang2025learning}. These promising results motivated the development of the POPSIM framework to further advance the SciML approach to data-driven simulation for operations and control.

\section{JAX Preliminaries}
JAX \cite{jax2018github} is a Python framework developed by Google for scalable numerical computing, with emphasis on Machine Learning (ML). We advise the reader to consult the documentation for a more detailed and up-to-date introduction to the framework, and restrict our coverage here to introducing concepts key to the rest of the paper.

At its core, JAX is about functional transformations of PyTrees. A PyTree is simply a tree of data that is registered with JAX. By default, standard Python containers such as dictionaries, lists, and tuples are supported, but custom data structures can also be easily registered. All JAX programs can be expressed as a pure function as such:
\begin{align}
    \mathbf{o} = f(\mathbf{u})
\end{align}
Where $\mathbf{u}$ is an input PyTree, $\mathbf{o}$ is an output PyTree, and $f$ is a pure Python function. Decorators can then be applied to the function to endow it with certain properties. For example:
\begin{enumerate}
    \item \texttt{jax.jit} just-in-time compiles the function.
    \item \texttt{jax.vmap} vectorizes the function (relevant for GPU parallelism).
    \item \texttt{jax.jacfwd}, \texttt{jax.jacrev}, \texttt{jax.grad} perform automatic differentiation (relevant for optimization and training).
\end{enumerate}
Libraries like Equinox \cite{kidger2021equinox} and Flax \cite{flax2020github} provide utilities to recover a limited set of Object-Oriented Programming (OOP). This is of particular use for defining machine learning models, which typically take the form of a parameterized function:
\begin{align}
    \mathbf{o} = f_\theta(\mathbf{u})
\end{align}
where $\theta$ is a tree of parameters. In code, an example is:
\begin{lstlisting}[language=Python]
import jax.numpy as jnp
import equinox as eqx

class Module(eqx.Module):
    array: jnp.array
    dict_of_arrays: dict[str, jnp.array]
    
    def __call__(self, inputs):
        # Implement the ``f" function.
        pass
\end{lstlisting}
Above, the set of parameters $\theta$ consists of both the array and dictionary of arrays. The user is free to implement whatever logic they would like in the call, whether that be neural networks, a physics formula, or a combination. The key requirement is that the code remains JAX compatible, which is a more detailed set of technical requirements best understood through the documentation.
\subsection{PyTree-Vector Equivalence}
\begin{figure}[!htb]
    \centering
    \includegraphics[width=0.9\linewidth]{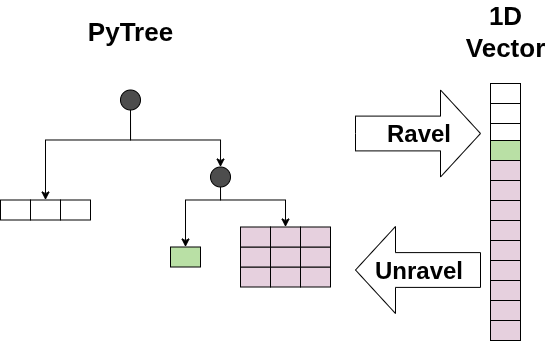}
    \caption{A depiction of the isomorphism between PyTrees and 1D vectors via the ravel operation and its inverse. Variables are assigned unique colors to depict an ordering in the raveled vector.}
    \label{fig:ravel_unravel}
\end{figure}

A concept of practical importance is the equivalence of PyTrees and 1D vectors. More precisely, PyTrees and 1D vectors in JAX are isomorphic under the ravel operation and its inverse, unravel. Figure \ref{fig:ravel_unravel} provides a depiction of this equivalence.

This equivalence is of particular use for developing data-driven simulators as principles-based models typically motivate data stored in a tree-like structure, while neural networks typically motivate a fixed vector or matrix input. ML workflows in fusion often involve boilerplate code to massage complex physics datasets into the right structure to be consumed by networks, but the capability to transform between a format friendly for physics and a format friendly for neural networks inside the ML model itself eliminates this step. 

Given this equivalence, we will refer to mathematical variables with vector notation, such as $\mathbf{x}$, as PyTrees as opposed to 1D vectors to emphasize the fact that support is provided for tree-like structures of data as well.

\section{Modules}
The POPSIM approach is to develop a collection of stand-alone simulation modules that can be hierarchically combined into larger modules. To facilitate this, POPSIM provides a collection of tools to:
\begin{itemize}
    \item Define simulation modules
    \item Run time-dependent simulations
    \item Parallelize modules across inputs and initial conditions
    \item Evaluate modules on datasets
    \item Train modules on datasets
    \item Track metrics and evaluations from training runs
    \item Export module parameters to mirrors on external software systems (e.g. control systems)
\end{itemize}
\subsection{Time-Independent Modules}
Time-independent modules are the simplest case, comprising only an input PyTree, $\mathbf{u}$, an output PyTree $\mathbf{o}$, and a PyTree of module parameters $\theta$:
\begin{align}
    \mathbf{o} = f_\theta(\mathbf{u})
\end{align}
Note that time-independent modules follow the exact same form as described in the section on JAX preliminaries.

\subsection{Time-Dependent Modules}
Time-dependent modules require special treatment, especially for control systems design. In POPSIM, time-dependent modules are defined as arbitrarily nonlinear state-space models. In continuous time, this is given by:
\begin{subequations}
\begin{align}
    \dot{\mathbf{x}} &= f_\theta(\mathbf{x}, \mathbf{u}) \\
    \mathbf{o} &= \mathcal{O}_\theta(\mathbf{x}, \mathbf{u})
\end{align}
\end{subequations}
Where the function $f$ is the dynamics function, which maps the state tree $\mathbf{x}$ and input tree $\mathbf{u}$ to the time derivative of state. The output function $\mathcal{O}$ maps state and inputs to additional module outputs (for example, by calculating derived quantities). The user may also specify discrete-time modules:
\begin{subequations}
\begin{align}
    \mathbf{x}_{t+1} &= f_\theta(\mathbf{x}_t, \mathbf{u}_t)\\
    \mathbf{o}_t &= \mathcal{O}_\theta(\mathbf{x}_t, \mathbf{u}_t)
\end{align}
POPSIM also provides support for specifying some subset of state variables as continuous time and some as discrete time. 
\end{subequations}
\subsection{Time-Stepping}
Time stepping can be accomplished with either Diffrax \cite{kidger2021on} or with a simple forward Euler method with a fixed time step. The former provides the advantages of support for irregular time-bases, a library of advanced numerical integration methods, and a library of adjoint methods for improved automatic differentiation. However, at time of writing, it does not provide support for discrete time variables, which motivates the implementation of the latter.
\subsection{Controllers as Time-Dependent Models}
Note that controllers can be viewed as time-dependent simulation modules. Consider the example of a discrete-time PID controller, which can be mathematically written as:
\begin{subequations}
\begin{align}
    I_{t+1} &= I_{t} + e_t\,\Delta t,\\
    D_{t} &= \frac{e_t - e_{t-1}}{\Delta t},\\
    a_t &= K_P\,e_t \;+\; K_I\,I_t \;+\; K_D\,D_t,
\end{align}
\end{subequations}
where $e_t$ is the error, $I_t$ is the accumulated integral state, $D_t$ is the derivative estimate, $\Delta t$ is the sampling interval, $a_t$ is the control action, and $K_P$, $K_I$, $K_D$ are control gains. This can be cast in the discrete-time module form, with the following notational re-mapping:
\begin{subequations}
\begin{align}
    \mathbf{x}_t\equiv \begin{bmatrix}I_t\\ e_{t-1}\end{bmatrix} &\quad \mathbf{u}_t\equiv [e_t] \quad\mathbf{o}_t\equiv [a_t]\quad \theta = \begin{bmatrix}K_P\\ K_I\\ K_D\end{bmatrix}\\
    &\underbrace{\mathbf{x}_{t+1} = \begin{bmatrix}
    I_t + e_t\Delta t\\
    e_t
    \end{bmatrix}}_{f_\theta}\\
    &\underbrace{a_t = K_Pe_t + K_II_t + K_DD_t}_{O_\theta}
\end{align}
\end{subequations}
The main conceptual shift from the simulation to control module case involves remapping inputs as controller references, outputs as control actions, and parameters as controller tunings.

\subsection{Hierarchical Modules}
As individual modules are developed, more complex modules can be built by hierarchically combining existing modules. To provide a very basic code example:
\begin{lstlisting}[language=Python]
class ModuleC(TimeIndepModule):
    moduleA: ModuleA
    moduleB: ModuleB

    def __call__(self, inputs):
        # Input module code.
        pass
\end{lstlisting}
Above, moduleA and moduleB are members of ModuleC. From the PyTree perspective, what is happening is the parameters of module C, $\theta_C$, consist of the concatenation of the parameters of modules A and B, $\theta_A$ and $\theta_B$. 

\subsection{Exporting Module Parameters}
At time of writing, the primary method of exporting modules to control systems is to convert the PyTree of module parameters $\theta$ into a JSON file via the \texttt{popsim.export} function, which also generates a netCDF4 file of expected module inputs and outputs. Deployment then involves re-implementing the logic of the module on the target control system, along with a helper function to load the JSON file. The validation data can then be used to validate that the implementation on the control system side is correct.

\section{Data Sources, Preparation, and Loading}
\subsection{Data Sources}
Tokamak operations is typically organized into ``pulses'', also known as ``discharges'' or ``shots'', which are events where instances of plasmas are created, sustained, and terminated. Each pulse involves multiple steps and contexts where data is generated. Prior to each pulse, user settings are provided to the control system, and pre-pulse calculations, and sometimes time-dependent simulations, are performed to provide informative data, such as control references, to the control system. During each pulse,  the data acquisition system collects time series data from a large suite of diagnostics, often at different time resolutions, which are fed into the plasma control system (PCS). The PCS then performs calculations using the collected data to make decisions in real-time that are sent to actuators. It is noteworthy that intermediate data generated by the PCS in the process of determining actuation commands is typically highly valuable for debugging purposes, and thus is often recorded as well. Since diagnostics typically can't directly observe the full state of the plasma, principles-based codes are typically run post-hoc to process the collected data into physically interpretable quantities. In addition, human operators typically make  observations about the experiment that are recorded in the logbook, which can be a rich source of important metadata, such as conditions in the facility at the time of the experiment, and qualitative assessments of the outcomes. In summary, each pulse generates six kinds of data:
\begin{enumerate}
    \item Pre-pulse control system settings set by the user
    \item Measurements made by real-time diagnostics
    \item Control system data, which includes both actuator commands and the product of intermediate calculations
    \item Pre-pulse analyses and simulations
    \item Post-pulse analyses and interpretative codes
    \item Logbook recordings of observations
\end{enumerate}
In today's fusion experiments, data types 1-3 are typically recorded using the MDSPlus system \cite{stillerman1997mdsplus}, with some of the data generated by pre-pulse and post-pulse codes recorded as well. However, data relevant to machine learning (ML) are often the product of additional codes that do not record data in MDSPlus. This is, in part, due to the fact that MDSPlus was designed prior to the age of ML and big data and, thus, can be relatively slow with handling large scale data queries. This limitation has motivated an ongoing project, MDSplusML \cite{stillerman2025mdsplusml}, which aims to bridge the gap with ML workflows. The IMAS data schema \cite{imbeaux2015design} aims to be the unifying schema with which fusion data is stored in the long run. However, at the time of writing, it has only recently been open sourced and broad adoption is still in the early stages.

Thus, in practice today, multiple heterogeneous data sources need to be handled. For example, the DEFUSE framework for post pulse data processing, analysis, and labeling \cite{pau2023modern}, and the Matlab Equilibrium Toolbox (MEQ) suite \cite{merle2024full} are two libraries in frequent use that generate data in their own custom formats which need to be handled. As discussed in the following subsections, the POPSIM strategy is to build tools to map data from multiple data-sources into the Xarray format.

\subsection{Choice of Xarray}
Xarray, which can be seen as a Python front-end for netCDF4 files, was chosen as the platform for data structures that POPSIM modules consume and generate. The choice of adoption was largely driven by the following considerations:
\begin{itemize}
    \item The Python API for IMAS, IMAS-Python \cite{imas-python}, has adopted Xarray and netCDF4 as a backend, making it a natural touchpoint as fusion datasets gradually migrate to the IMAS schema.
    \item It was developed by the geoscience community, as a part of the Pangeo project \cite{eynard2019pangeo}, for the purpose of handling labeled spatial data, providing a large volume of built-in solutions for common data analysis and processing tasks. Existing solutions have been demonstrated to work from laptop scale to petabyte scale datasets on the cloud.
    \item Existing tools such as Xbatcher \cite{Jones_xbatcher_2024} help enable integration with machine learning models.
    \item The GraphCast \cite{lam2023learning} weather prediction model developed a preliminary integration of Xarray objects with JAX, by registering them as JAX Pytrees. This implementation inspired the xarray\_jax Python package which enables simulation modules to directly operate on Xarray objects, enabling direct integration of simulation modules with datasets.
\end{itemize}

\subsection{Building Tensorized Multi-Episode Datasets}
One of the challenges of applying machine-learning techniques to fusion datasets is the need to combine a large number of heterogeneous trees of data into a single tensorized tree. Major fusion experiments typically have $\mathord{\sim}10^4$ to $\mathord{\sim}10^5$ pulses throughout their often multi-decade lifespans, and additional trees of simulated pulses are also of relevance. In POPSIM, we adopt the term ``episode'' from the reinforcement learning (RL) literature as a generic term encompassing both experimental and simulated pulses.

\begin{figure*}[!htbp]
    \centering
    \includegraphics[width=\textwidth]{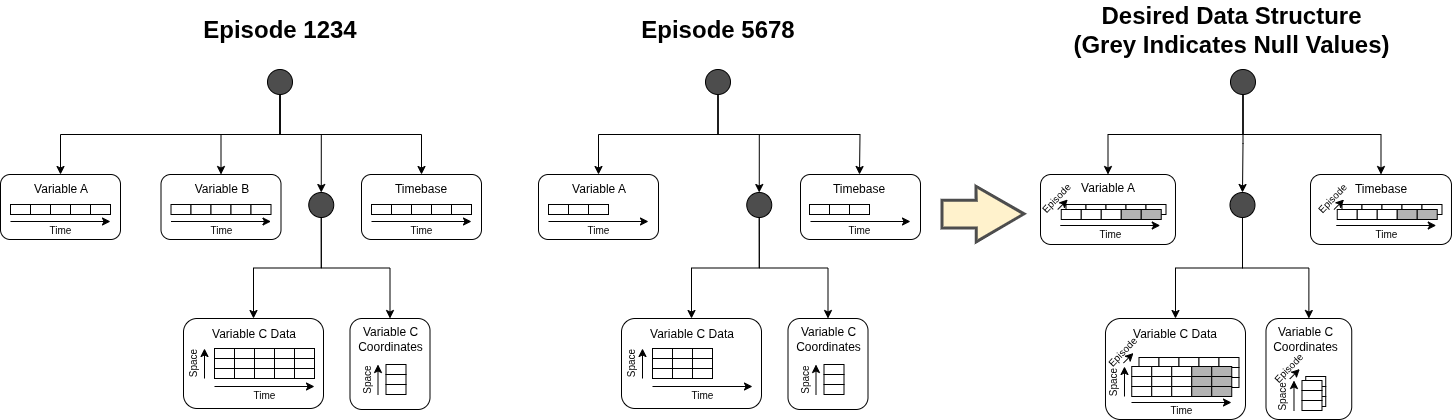}
    \caption{Diagram depicting the process of building a multi-episode tensorized dataset out of multiple heterogeneous trees of data. Note that one common challenge is that trees often have different time bases. This is handled in this case by simple making time an additional variable. Grey blocks in the resulting data structure represent padding.}
    \label{fig:build-ds}
\end{figure*}

The POPSIM approach is to provide a helper function \texttt{build\_tensorized\_dataset} which does the job of building this tensorized dataset, given a data source specific function that processes a single tree. The process is depicted in Figure \ref{fig:build-ds}. Because the resulting dataset may be larger than memory, we leverage the Zarr \cite{zarr-python-3.1.0} storage format, which allows us to incrementally build up a dataset that does not fit into memory.

\begin{figure}[!htb]
    \centering
    \includegraphics[width=\linewidth]{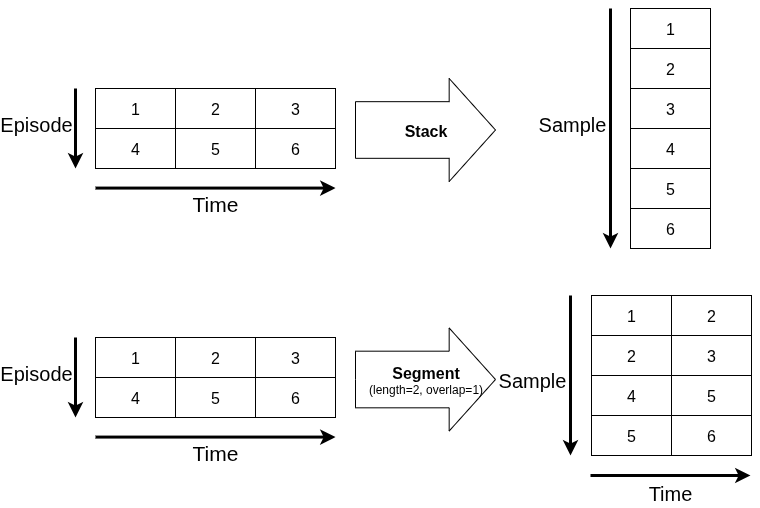}
    \caption{Example depiction of the stack and segment operations. Note that the segment operation is done with a length of two and overlap of one, resulting in samples that share a single element.}
    \label{fig:stack-segment}
\end{figure}

\subsection{Dataloaders}
POPSIM implements standard dataloaders that feed data in \texttt{xr.Dataset} format into modules. Optional flags are provided to support standard features such as batching, shuffling, and splitting of datasets by episode. In addition, a helper function \texttt{make\_standard\_dataloaders} is provided which generates training, validation, and test dataloaders with shuffling applied to the training dataloader only.

A distinction is made between time-independent dataloaders and time-dependent dataloaders. In the time-independent case, GPU parallelism can be applied across all time slices for all episodes. To achieve this, the episode and time dimensions are stacked to form a new sample dimension. The time-dependent dataloader provides the option to segment episodes into smaller episodes, with the option to have segments overlap each other. This is motivated by two considerations: 1) training is faster with shorter segments as GPUs are comparatively slow at sequential operations, and 2) shorter sequences help ameliorate the vanishing gradient problem \cite{hochreiter1998vanishing}. Figure \ref{fig:stack-segment} depicts the stack and segment operations, which produce batches with a sample dimension. JAX vectorization is then applied to run the module in parallel across rows of the sample dimension.

\section{Module Simulation, Evaluation, and Training}
\subsection{Simulation and Evaluation}
The POPSIM framework is designed to have common functions perform the job of simulating individual modules. There are three primary APIs for executing modules, motivated by different use cases, described below.

\texttt{popsim.simulate.simulate} is the API for users who wish to manually specify simulation intial states and time-dependent inputs. It takes in a time-dependent module, a time base, initial state(s), time-dependent input(s), and performs the full time-dependent simulation for the full time base. The API supports performing batches of simulation, which are expressed by providing lists of initial states and time-dependent inputs. Additional utilities are provided in the package for generating such batches of simulation inputs. The output is a \texttt{xr.Dataset} containing simulation data for all simulation cases.

\texttt{popsim.simulate.single\_step} is the API developed for performing closed-loop simulation against external pieces of software (such as control systems). It performs a single time step of a module, given a state PyTree $\textbf{x}_t$ and an input PyTree $\textbf{u}_t$ for a single time step. The output is an output PyTree $\mathbf{o}_t$ and the next state $\mathbf{x}_{t+1}$.

\texttt{popsim.eval\_on\_data} is the API developed for machine learning style workflows, where the module is run using data from a dataset, and metrics and summary figures are reported as opposed to user-specified manual inputs. As such, it takes as input a simulation module along with a dataloader that provides all of the data needed to run the module. By default, it outputs an $\texttt{EvalData}$ named tuple which contains the module, input dataloader, and output dataset. However, the user can provide an $\texttt{EvaluationSuite}$, which is a dictionary of functions that take in $\texttt{EvalData}$ and generate figures and metrics for the model. 
\subsection{Module Training}
Module training infrastructure, inspired in part by PyTorch Lightning \cite{Falcon_PyTorch_Lightning_2019}, is provided under the \texttt{popsim.ml} namespace, which provides two primary levels of entry points.

\texttt{popsim.ml.Trainer}. One option is for the user to manually define the code required to build a \texttt{Trainer} instance calling the \texttt{train} method attached. This involves:
\begin{itemize}
    \item Instantiating the module
    \item Instantiating dataloaders
    \item Instantiating an optimizer
    \item Defining a loss function
    \item (Optional): instantiating evaluation suites and loggers.
\end{itemize}
Doing so provides the user with more fine-grained control over the training run. However, taking the lower level approach leads to a considerable amount of duplicate code, and, more importantly, does not strictly require management of training run configuration, which is necessary for both hyper-parameter sweeps and proper tracking of experiments.

\texttt{popsim.ml.launch}. This higher level abstraction requires the user to define a \texttt{TrainingRunBuilder} for their specific module. The purpose of the \texttt{TrainingRunBuilder} is to map a configuration dictionary, stored either as a Python dictionary or a yaml, to the set of objects necessary to automatically perform the training run. Once a \texttt{TrainingRunBuilder} has been defined, the user can use the \texttt{popsim.ml.launch} interface to launch a training run given a config dictionary and a \texttt{TrainingRunBuilder}.

\subsection{Experiment Tracking and Hyperparameter Sweeps}
By default, POPSIM integrates with Weights and Biases (W\&B) \cite{wandb} for the purposes of logging training metrics, tracking experiments, and launching hyper-parameter sweeps. In addition to launching training runs, the \texttt{popsim.ml.launch} API provides utilities for launching W\&B hyper-parameter sweeps and agents that manage training runs.

\section{Example: Time-Dependent Transport Prediction with TCV Data}
As an illustrative example, we report on a simulation module developed in POPSIM that predicts the time-dependent transport dynamics of TCV plasmas, using only engineering parameters as inputs, similar to models developed for prior works \cite{boyer2021prediction, wang2025learning, boyer2020toward}. A more detailed report of the architecture, results, statistics, and implications will be reported in subsequent works.

The purposes of the module is to predict the time evolution of the plasma stored energy, $W_{tot}$, and kinetic profiles as measured by Thomson Scattering: $T_e$ and $n_e$, which are the electron temperature and density, respectively, and are functions of the normalized radius of the plasma, $\rho$. The inputs are restricted to the standard engineering parameters of plasma current, $I_p$, minor radius, $a_{minor}$, elongation, $\kappa$, upper and lower triangularity, $\delta_u$ and $\delta_l$, auxiliary heating, $P_{aux}$, and average electron density $\bar{n}_e$. The state, inputs, and outputs are defined as such:
\begin{subequations}
    \begin{align}
    \mathbf{x} = [W_{tot}]\quad \mathbf{u} &= [I_p, a_{minor}, \kappa, \delta_u, \delta_l, P_{aux}, \bar{n}_e]\\
    \mathbf{o} &= [W_{tot}, T_e(\rho), n_e(\rho)]
\end{align}
\end{subequations}
We note that all input parameters are highly controllable quantities, with the exception of the line-averaged density, which future work will aim to predict as well in response to gas valves and other control variables. The module involves four submodules for predicting ohmic heating, radiated power, energy confinement time, and profiles.

\begin{figure}
    \centering
    \includegraphics[width=0.9\linewidth]{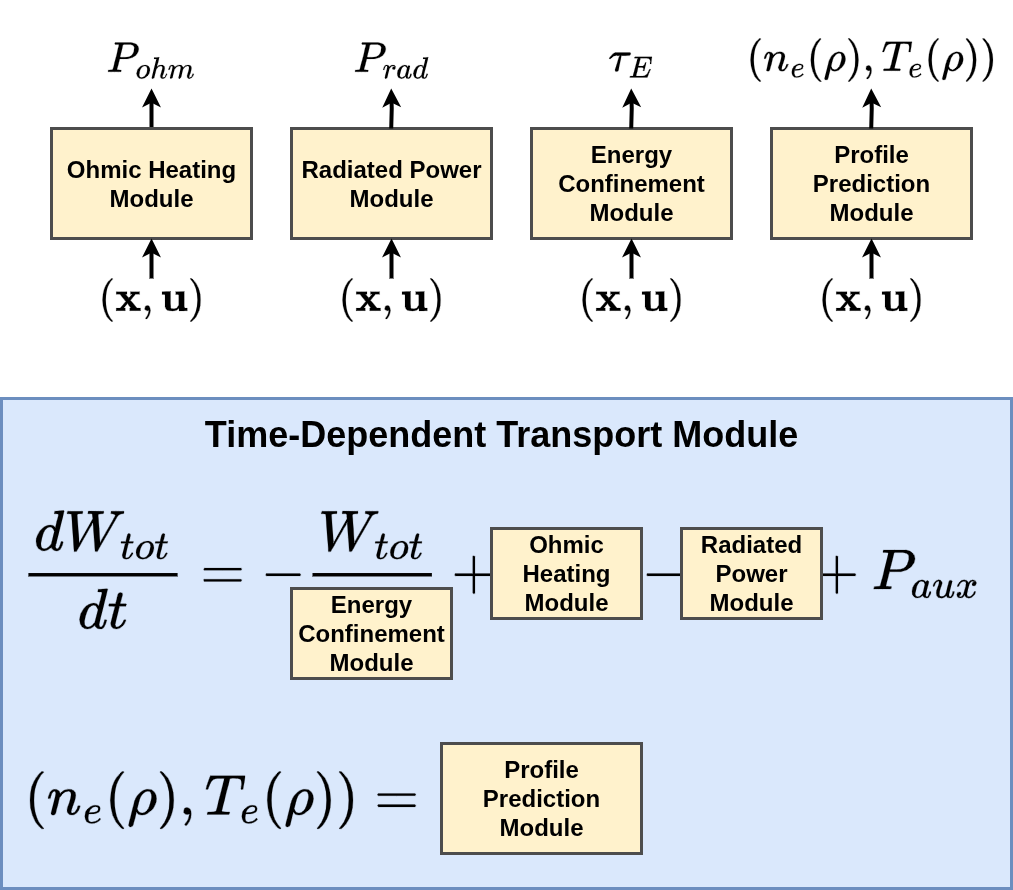}
    \caption{Block diagram depiction of the time-dependent transport module, which consists of four time-independent submodules.}
    \label{fig:transport-module-block}
\end{figure}

Figure \ref{fig:transport-module-block} shows the architecture of the module. The ohmic heating, radiated power, and profile prediction modules are first trained in a time-independent fashion to predict their respective quantities. These modules are then integrated into the time-dependent transport module, where the parameters of all the submodules are further trained as part of the time dependent module. That is, forward simulation is done to yield predictions, these predictions are then compared to experimental data, and automatic differentiation is applied to find the gradients to update the module to improve the time-dependent predictions.

\begin{figure}
    \centering
    \includegraphics[width=0.9\linewidth]{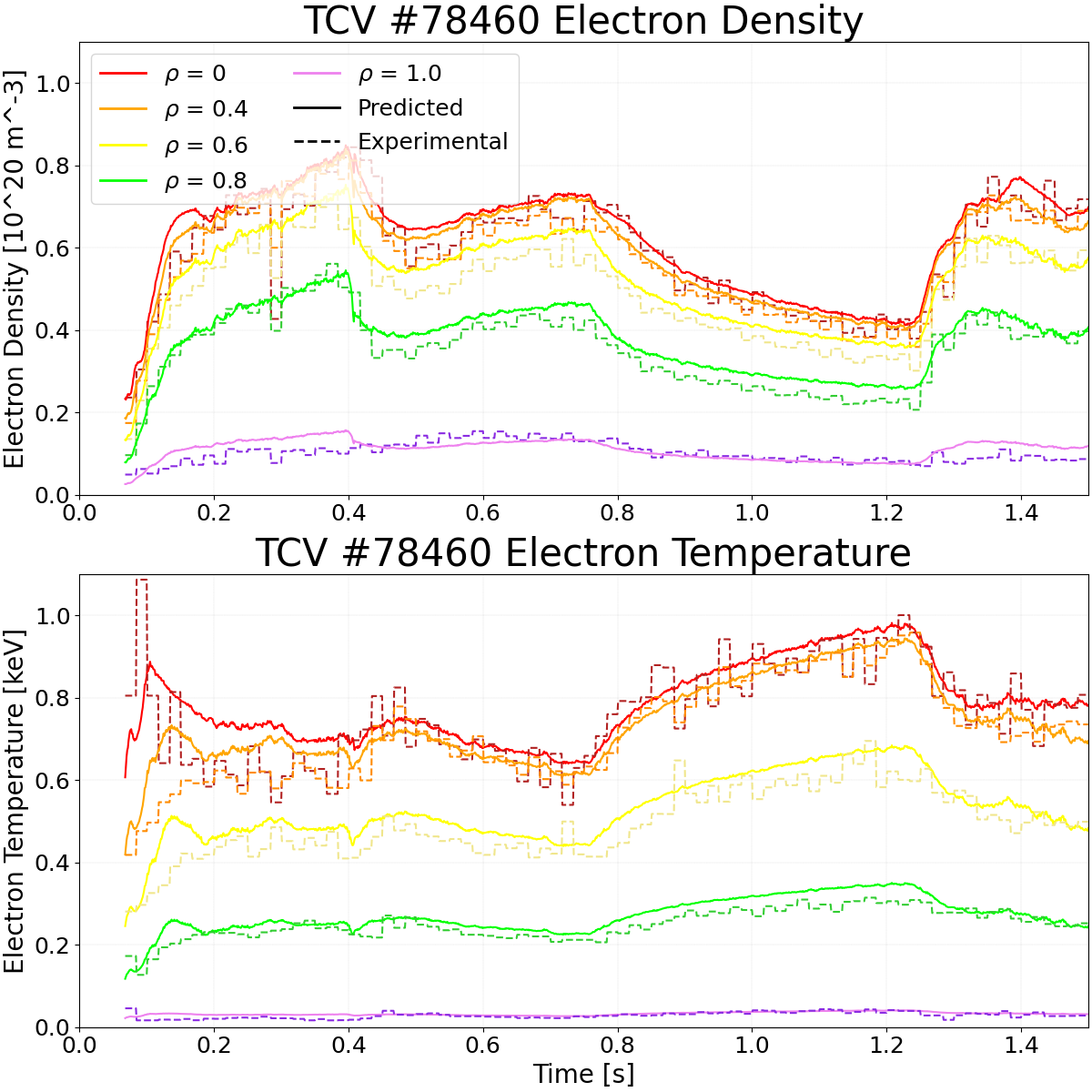}
    \caption{Example experimental measurements for a TCV pulse, along with predictions made by the time-dependent model for the full pulse.}
    \label{fig:tcv-transport-example}
\end{figure}

\section{Example: Time-Dependent POPCON}
\begin{figure}[!htbp]
    \centering
    \includegraphics[width=1.0\linewidth]{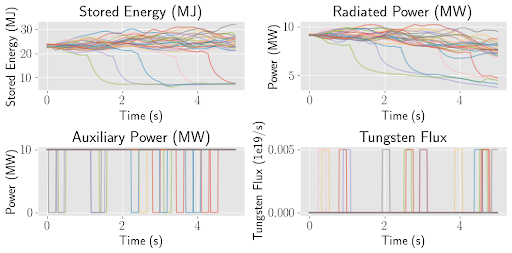}
    \caption{Example simulation results from the time-dependent POPCON module, showing the effect of impurity injection and loss of auxiliary heating events.}
    \label{fig:td-popcon}
\end{figure}

Plasma operational contour (POPCON) codes, such as CFSPOPCON \cite{body2023sparc}, are often used for low fidelity rapid scoping of steady-state plasma scenarios. Leveraging the CFSPOPCON library as a dependency in POPSIM, we develop a time-dependent POPCON module that enables rapid, massively parallel simulation of off-normal events. Figure \ref{fig:td-popcon} shows example simulations of off-normal events from this module.

\section{Conclusion}
This paper reported on the key technical architecture of the POPSIM framework, which is currently under development to address simulation challenges encountered in tokamak operations and control. High level technical functionalities of the modules system, dataset preparation, and training and evaluation pipelines are specified. Illustrative examples from a time-dependent transport module trained on TCV data and a time-dependent POPCON are shown. Future works will report in further detail modeling results on TCV data. In addition, the code repository is planned to be open-sourced in the near future.

\section*{Acknowledgments}
This work was funded in part by Commonwealth Fusion Systems (CFS).

\bibliographystyle{IEEEtran}
\bibliography{references}

\end{document}